# Photonic analog computing with integrated silicon waveguides

**Jianji Dong, Ting Yang, Aoling Zheng, and Xinliang Zhang**

*The spectra of silicon integrated waveguides are tailored to process analog computing (i.e., differential and integral) in optical domain with huge bandwidth.*

Photonic analog computing is used to process optical computing (i.e., differential and integral) of analog signals without analog-to-digital conversion (ADC), digital-to-analog conversion (DAC) or optical-electrical-optical (OEO) conversion, which offers huge bandwidth of optics. There are two fundamental computing systems of analog computing, differentiator and differential equation solver. Photonic differentiator (DIFF) has wide applications in numerous fields such as pulse characterization, ultra-fast signal generation, and ultra-high-speed coding. And the optical differential equation (ODE) system can be used in many field of science and engineering, such as temperature diffusion processes, opto-mechanical system, and the response of different resistor-capacitor photonic circuits, etc.

Photonic DIFFs and ODE solver systems have been demonstrated by many approaches. For example, the photonic DIFFs can be implemented with semiconductor optical amplifiers (SOAs) [1, 2] and highly nonlinear fibers [3]. The ODE solvers were implemented by an optical feedback loop [4]. Nevertheless, the configurations of all these schemes are bulky, complex and high power consumption. Since photonic integrated circuits for photonic computing open up the possibility for the realization of ultrahigh-speed and ultra wide-band signal processing with compact size and low power consumption, we focus on the solution with integrated silicon photonics. The photonic analog computing can be regarded as a linear system. Thus the spectra of silicon integrated waveguides should be tailored to meet the requirements of analog computing. Here we demonstrate several DIFFs and ODE solvers with integrated silicon Mach-Zehnder Interferometer (MZI) and microring resonator (MRR), respectively. Figures 1(a) and 1(b) illustrate typical examples of photonic DIFF with silicon MZI and ODE solver with MRR, respectively.

For example, to implement *N*th-order photonic DIFF, the corresponding transfer function should be designed with $H(\omega) = (j\omega)^N$. We proved that the MZI structure has a good linear frequency response near the MZI resonant notch, acting as the first-order DIFF [5]. Therefore, to achieve *N*th-order DIFF, the MZI unit just needs to be cascaded by *N* times with the resonant frequencies aligned. We design and fabricate cascaded MZIs on commercial silicon-on-insulator (SOI) wafer. Figure 2(a) shows the microscope image of our on-chip MZIs. Different MZI units (i.e., MZI-1, MZI-2, MZI-3) are used to implement different order DIFF. Then, a Gaussian pulse train, shown in Figure 2(b), is injected into the MZI unit. We measure the output temporal waveforms of the first-order, second-order, and third-order DIFFs, as shown in Figures 2(c)-(e), respectively. It can be seen that the measured differentiated pulses fit well with the simulated pulses.

In fact, when the order number *N* is a fractional number, the DIFF can be named as fractional-order DIFF. Fractional-order DIFF can be considered as a generalization of integer-order DIFF, with potentials to accomplish what integer-order DIFF cannot. We notice that the phase jump of frequency response of MZI unit is not exactly π, if the powers of two arms are not equal. Hence, we can tune the loss of one arm by carrier modulation with external power supply, resulting in a variable phase jump of MZI response. Therefore, the MZI unit can be used to implement fractional-order DIFF. We then fabricate the MZI unit with PN junction at the commercial 0.18µm complementary metal-oxide semiconductor (CMOS) foundry [6]. Figure 3 shows the micrographs of (a) total MZI structure, (b) coupling grating, (c) multimode interferometer (MMI), and (d) integrated *p-i-n* diode. A Gaussian-like pulse train is launched into the chip sample, as shown in Figure 3(e). When we apply different voltages on the MZI arms, we measure the output differentiated waveforms, as shown in Figures 3(f)-3(j). At the same time, the simulated waveforms of fractional-order DIFF are also shown with a perfect accordance, whose fractional orders are *N* = 0.83, 0.85, 0.98, 1.00, and 1.03, respectively.

Despite of the silicon MZI units, we also fabricate silicon MRR to demonstrate high-order DIFF and fractional-order DIFF, respectively [7]. The difference is that the MRR normally has a much smaller bandwidth than that of MZI, so MRR is more suitable for processing of narrow-bandwidth analog signals.

Another important application of analog computing is ODE solver. The transfer function should be designed with $H(\omega) = \dfrac{1}{j\omega + k}$ to solve an ODE. Interestingly, we find that the frequency response at drop port of an add-drop MRR inherently matches with that of the first-order ODE solver. Furthermore, if we can change the *Q* factor of the MRR, we can make the constant-coefficient of ODE solver tunable [8]. Figures 4 (a) and (b) show the microscope images of the fabricated MRR and the zoom-in ring region, respectively. The ring waveguide was controlled by the electrodes so that the *Q* factor of ring can be changed by the applied voltages. We choose super-Gaussian pulse as the input signal, as shown in Figure 4(c). When the voltage is 0V and 1.3V, the output waveforms are shown in Figures 4(d) and 4(e), respectively. We calculate the constant-coefficient, which is 0.038/ps and 0.082/ps, respectively. The calculated waveforms accord well with the measured ones. Thus the ODE solver is very accurate within certain bandwidth.

With the theory of signal and system, we design some silicon integrated devices (MZI and MRR) to implement photonic DIFF and ODE solver. The basic principle is to tailor the spectra of silicon integrated waveguides to meet the requirements of analog computing circuits. These analog photonic integrated circuits are very promising in future computing systems with high speed, low cost, and compact size. We also plan to employ these basic computing units in more complex computing modules.

**Jianji Dong, Ting Yang, Aoling Zheng, and Xinliang Zhang**

Huazhong University of Science and Technology

Wuhan, China

Jianji Dong received his PhD in 2008 and is currently a professor in Wuhan National Lab for Optoelectronics. His research interests include photonic digital/analog computing, optical vortex manipulation, and microwave photonics. He has published more than 90 peer-reviewed journal papers.

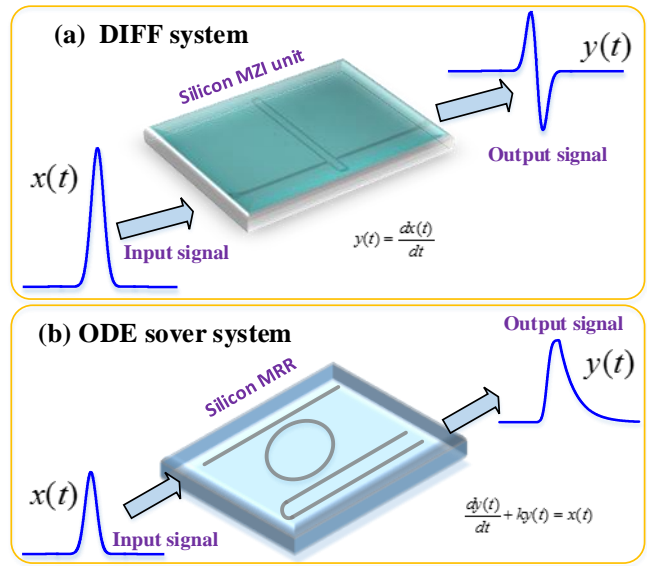

Figure 1 Illustration of (a) Photonic DIFF system with a silicon MZI unit, and (b) ODE solver system with a silicon MRR.

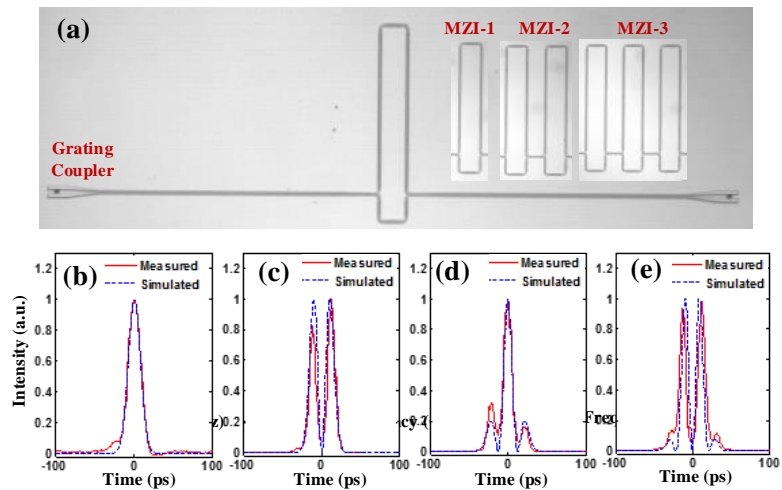

Figure 2 (a) Microscope image of our on-chip MZIs, insets: photos of structures of MZI-1, MZI-2, and MZI-3, (b) input pulse, (c)-(e) temporal waveforms for 1st-, 2nd-, and 3rd-order DIFFs, respectively

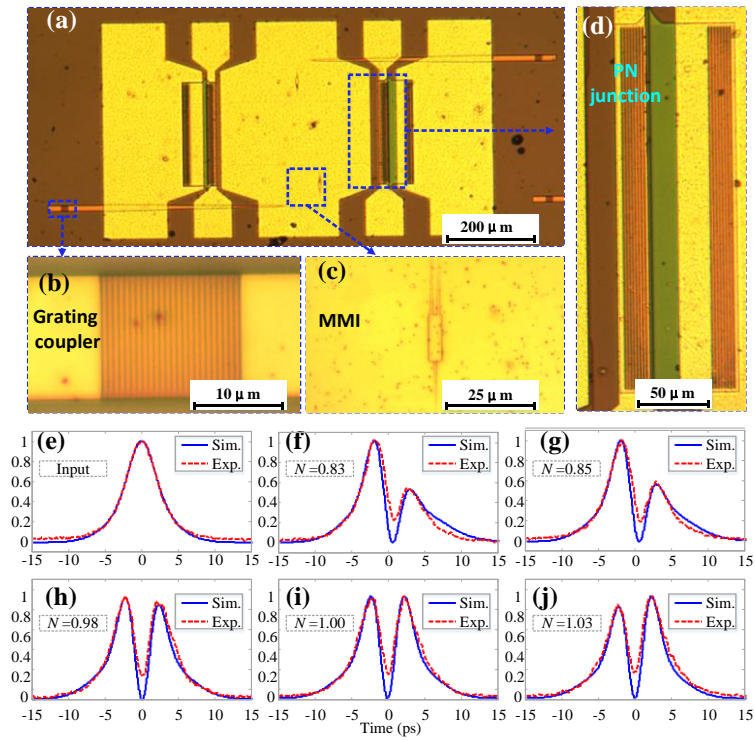

Figure 3 (a) Micrographs of MZI, (b) coupling grating, (c) MMI, (d) integrated *p-i-n* diode, (e) An input Gaussian-like pulse, (f)-(j) output fractional-order DFF waveforms with order number N = 0.83, N =0.85, N =0.98, N =1.00, N = 1.03, respectively.

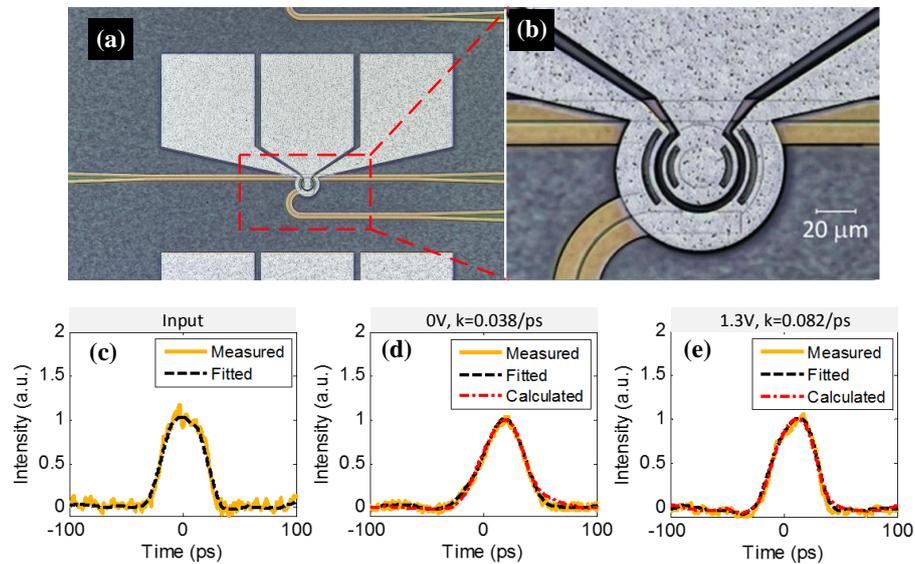

Figure 4 (a) Microscope image of the fabricated MRR, (b) microscope image of the zoom-in ring region, (c) the input super-Gaussian waveform, (d) and (e) measured output waveforms with different voltages applied, yellow solid line: measured waveforms, black dash line: fitted waveforms, and red attunement line: calculated ideal outputs.